\newcommand{\nustar}{{\sl NuSTAR}\xspace}

\newcommand{\xrt}{{\sl Swift}/XRT\xspace}
\newcommand{\fermi}{{\sl Fermi}\xspace}
\newcommand{\xmm}{{\sl XMM-Newton}\xspace}

\documentclass{aa}  
\newcommand{\fgl}{4FGL J1405.1-6119\xspace}
\usepackage{graphicx}	
\usepackage{amsmath}	
\usepackage{hyperref}
\usepackage{subcaption}
\usepackage{ulem}
\usepackage{txfonts}
\usepackage[]{hyperref}
\hypersetup{unicode=true, colorlinks=true, linkcolor=[rgb]{0.53, 0.15, 0.34}, citecolor=[rgb]{0.0, 0.27, 0.42}, filecolor=[rgb]{1.0, 0.13, 0.32}, urlcolor=[rgb]{0.53, 0.15, 0.34}}
 
\begin{document}

   \title{\textit{NuSTAR} and \textit{XMM-Newton} observations of the binary 4FGL J1405.1-6119}

   \subtitle{A $\gamma$-ray emitting microquasar?}

\author{Enzo A. Saavedra\inst{1},
       Federico A. Fogantini\inst{2},
        Gast\'on J. Escobar\inst{3,4}, 
        Gustavo E. Romero\inst{1,2},  \\
        Jorge A. Combi\inst{1,2,5} \and
        Estefania Marcel\inst{1}}

   \institute{Facultad de Ciencias Astron\'omicas y Geof\'{\i}sicas, Universidad Nacional de La Plata, Paseo del Bosque, B1900FWA La Plata, Argentina \and 
   Instituto Argentino de Radioastronom\'ia (CCT La Plata, CONICET; CICPBA; UNLP), C.C.5, (1894) Villa Elisa, Buenos Aires, Argentina \and
   Physics and Astronomy Department Galileo Galilei, University of Padova, Vicolo dell'Osservatorio 3, I-35122, Padova, Italy \and
   INFN - Padova, Via Marzolo 8, I--35131 Padova, Italy \and  
   Departamento de F\'\i sica (EPS), Universidad de Ja\'en, Campus Las Lagunillas s/n, A3, 23071 Ja\'en, Spain}

   \date{Received; accepted}
 
  \abstract
   {\fgl  is a high-mass $\gamma$-ray-emitting binary that has been studied at several wavelengths. The nature of this type of binary is still under debate, with three possible scenarios usually invoked to explain the origin of the $\gamma$-ray emission: collisions between the winds of a rapidly rotating neutron star and its companion, collisions between the winds of two massive stars, and nonthermal emission from the jet of a microquasar.}
   {We analyzed two pairs of simultaneous {\it NuSTAR} and \textit{XMM}-{\it Newton} observations to investigate the origin of the radio, X-ray, and $\gamma$-ray emissions. }
   {We extracted light curves between 0.5 and 78 keV from two different epochs, wich we call Epoch 1 and Epoch 2. We then extracted and analyzed the associated spectra to gain insight into the characteristics of the emission in each epoch. To explain these observations, along with the overall spectral energy distribution, we developed a model of a microquasar jet. This allowed us to make some inferences about the origin of the observed emission and to discuss the nature of the system.}
   {A power-law model combined with the inclusion of a blackbody accurately characterizes the X-ray spectrum. The power-law index ($E^{-\Gamma}$) was found to be $\sim$ 1.7 for Epoch 1 and $\sim$ 1.4 for Epoch 2. Furthermore, the associated blackbody temperature was $\sim$ 1 keV and with a modeled emitting region of size ${\lesssim}~16$ km. The scenario we propose to explain the observations involves a parabolic, mildly relativistic, lepto-hadronic jet. This jet has a compact acceleration region that injects a hard spectrum of relativistic particles. The dominant nonthermal emission processes include synchrotron radiation of electrons, inverse Compton scattering of photons from the stellar radiation field, and the decay of neutral pions resulting from inelastic proton-proton collisions within the bulk matter of the jet. These estimates are in accordance with the values of a super-Eddington lepto-hadronic jet scenario. The compact object could be either a black hole or a neutron star with a weak magnetic field. Most of the X-ray emission from the disk could be absorbed by the dense wind that is ejected from the same disk.}
   {We conclude that the binary \fgl could be a supercritical microquasar similar to SS433.}

   \keywords{X-rays: binaries -- gamma-rays: stars -- stars: individual (CXOU J053600.0-673507, 4FGL J1405.1-6119) 
               }

   \titlerunning{A $\gamma$-ray emitting microquasar?}
   \authorrunning{Enzo A. Saavedra et al.}

   \maketitle
%

\section{Introduction} \label{sec:intro}

Binary sources containing neutron stars (NSs) or black holes (BHs) dominate the Galactic X-ray emission above 2 keV \citep[see, e.g.,][]{Grimm2002A&A...391..923G}. These systems are called X-ray binaries and are usually divided into two major classes, high-mass X-ray binaries and low-mass X-ray binaries, according to the mass of the donor star (mass $\gtrsim 8~M_{\odot}$ for the former and $\lesssim 8~M_{\odot}$ for the latter).

Within the high-mass X-ray binary class, three types of non-transient systems can emit $\gamma$-ray radiation \citep{Chernyakova2019A&A, ChernyakovaandMalyshev2020}: 

1. {\sl Colliding-wind binaries}   involve the interaction of two massive stars, whose nonrelativistic winds collide and produce $\gamma$-ray emission. Prominent examples of such systems are \object{$\eta$-Carinae}, \object{WR11}, and \object{Apep}. In colliding-wind binaries, intense shocks occur in the wind collision region, leading to the formation of a very hot plasma (> $10^6$ K). In addition, these systems have the ability to accelerate relativistic particles \citep{Eichler1993, Benaglia2003A&A}, which classifies them as particle-accelerating colliding-wind binaries \citep[e.g.,][and references therein]{DeBecker2013A&A, delpalacio2023A&A}.

\begin{figure*} [h!] \centering
    \includegraphics[width=\textwidth]{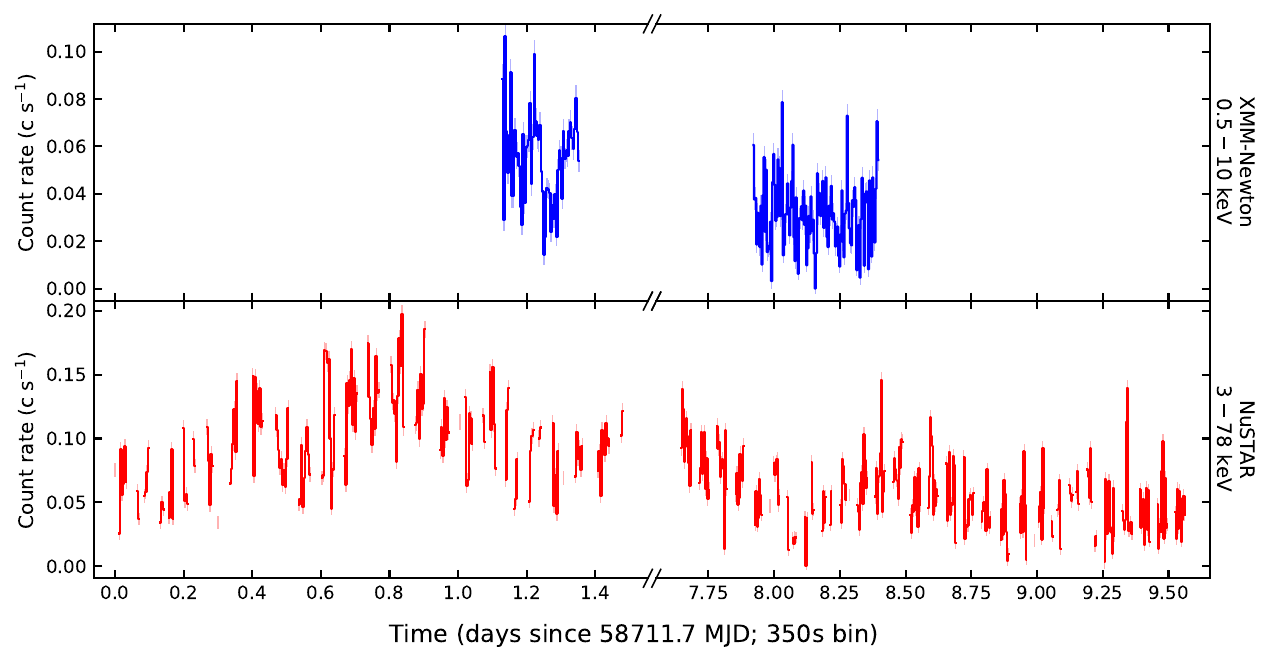}
    \caption{Background-corrected light curves of \fgl observed by {\sl NuSTAR} (FPMA+B; red) and \textit{XMM}-{\sl Newton} (EPIC pn+MOS; blue) with a binning of 350s, starting at 58711.705573~MJD. The first section of the light curve corresponds to the observation from August 18 (Epoch 1), while the second light curve is associated with the observation from August 25 (Epoch 2).} 
    \label{fig:lcnustar}
\end{figure*}

2. {\sl Gamma-ray binaries} are characterized by the presence of a young, magnetized, and rapidly rotating NS that emits a relativistic wind. This wind collides with the nonrelativistic wind of the companion OB star, and through this interaction they emit nonthermal radiation that peaks at high energies ($E>100$~MeV). 
These systems are thought to represent a short phase in the evolution of massive binaries, which comes after the birth of the compact object (CO) and is followed by the X-ray binary phase, in which the CO accretes matter from its companion \citep[e.g.,][and references therein]{Dubus2017, saavedraLSImnras}. In the latter phase, the nonthermal emission of the system peaks in X-rays. 
Examples of $\gamma$-ray binaries are \object{LS I+61$^{\circ}$303}, \object{PSR J1259-63}, \object{PSR J2032+4127}, \object{LS 5039}, \object{1FGL J1018.6-5856}, \object{and HESS J0632+057}.

3. {\sl Microquasars} (MQs) differ from the previous two categories in that their emission of $\gamma$-rays does not come from wind collisions. Instead, it comes from the jets ejected by the CO (BH or NS) and their interaction with the environment. Examples of MQs are \object{Cyg X-1}, \object{Cyg X-3}, and \object{SS433}

In total, there are nine known non-transient $\gamma$-ray-emitting binaries in our Galaxy \citep[e.g.,][and references therein]{Corbet2019ApJ...884...93C, Dubus2017, ChernyakovaandMalyshev2020}. Five of these systems contain an O-type star: \object{LS 5039}, \object{LMC P3}, \object{1FGL J1018.6-5856}, \object{HESS J1832-093}, and \object{4FGL J1405.1-6119}. The remaining four contain a Be star: \object{HESS J0632+057}, \object{LS I+61$^{\circ}$303}, \object{PSR B1259-63}, and \object{PSR J2032+4127}. Radio pulsations were observed in three of these systems -- \object{PSR J2032+4127} (\citealt{Camilo2009}), \object{PSR B1259-63} (\citealt{Moldon2014}), and \object{LS I+61$^{\circ}$303} (\citealt{Weng2022NatAs...6..698W}) -- evidence in favor of the presence of a NS. 
Many more galactic $\gamma$-ray binaries are expected to exist. We recall that the nature of many of the $\gamma$-ray sources detected by telescopes such as {\it Fermi}-LAT is not known, and some of the sources may be $\gamma$-ray binaries. In fact, the number of detectable Galactic $\gamma$-ray binaries is estimated to be $\sim 100$ \citep{Dubus2017}.

\fgl (also known as 3FGL J1405.4-6119) is a high-mass $\gamma$-ray-emitting binary first studied by \citet{Corbet2019ApJ...884...93C}. Using \fermi and \xrt data, \citet{Corbet2019ApJ...884...93C} found a strong modulation of $13.713~\pm~0.002$ days associated with the system's orbital period. The companion is classified as an O6.5 III star with a mass of about $25-35$~M$_{\odot}$ \citep{Mahy2015A&A...577A..23M}. The absence of partial and total eclipses suggests that this system has a low inclination ($60~\degr$).

To explain the origin of its $\gamma$-ray emission, \citet{Corbet2019ApJ...884...93C} proposed that \fgl~may be a $\gamma$-ray binary.  This hypothesis draws on analogies with other similar systems, such as \object{1FGL J1018.6-5856} and \object{LMC P3}. \citet{Xingxing2020} modeled the GeV time behavior of \fgl in the $\gamma$-ray binary scenario, assuming a binary consisting of a young pulsar and an O-type main sequence star. Conversely, the radio (5.5 GHz and 9 GHz) and X-ray (0.2--10 keV) luminosities show a positive correlation \citep{Corbet2019ApJ...884...93C}, as expected in MQs \citep{Falcke2004A&A}. 

We were able to perform a detailed temporal and spectral study of \fgl over a broad energy range by analyzing simultaneous \xmm and \nustar observations. In this paper we present our results and conclusions about this source. 
In \hyperref[sec:data]{Sect.~\ref{sec:data}} we present the X-ray observations and the corresponding tools used for their analysis. Our main results are presented in \hyperref[sec:results]{Sect.~\ref{sec:results}}. A jet model for the source is introduced and discussed in \hyperref[sec:results]{Sect.~\ref{sec:model}}. We discuss our results and present our conclusions in \hyperref[sec:disc]{Sect.~\ref{sec:disc}}.

\section{Observation and Data Analysis} \label{sec:data}

 \subsection{\xmm data}
 
The \xmm\  observatory is equipped with an optical instrument and two X-ray instruments: the Optical Monitor, which is mounted on the mirror support platform and provides coverage between 170 nm and 650 nm of the central 17 arcminute square region; the European Photon Imaging Camera (EPIC); and the Reflecting Grating Spectrometers (RGS). The EPIC instrument comprises three detectors -- the pn camera \citep{Struder2001A&A...365L..18S} and two MOS cameras \citep{Turner2001A&A...365L..27T} -- which are most sensitive in the $0.3-10$~keV energy range. The RGS instrument comprises two high-resolution spectrographic detectors sensitive in the energy range $0.3-2$~keV.

\xmm observed \fgl on August 17, 2019, with an exposure time of 32 ks (ObsID 0852020101) and on August 24, 2019, with an exposure time of 44 ks (ObsID 0852020201). In both observations, the MOS cameras were in large window mode, and the pn camera was in timing mode. 

We reduced the \xmm data using Science Analysis System (SAS)~v20.0 and the latest available calibration in early 2022. To process the observation data files, we used the \texttt{\footnotesize EPPROC} and \texttt{\footnotesize EMPROC} tasks. 
We selected circular regions with radii of 18 arcsec and 36 arcsec for the source and the background, respectively, with the latter away from any source contamination.
We then filtered the raw events lists, removing the high-energy, single-pattern particle backgrounds and thus creating cleaned event lists for each camera and observation. The resulting exposure times after background filtering are 19~ks (59\% of total) for the first observation, and 44~ks (100\% of total) for the second observation.
We studied the presence of pile-up with the {\tt {\footnotesize EPATPLOT}} task and did not find any deviation of the data from the expected models; thus, no excision radii were applied.  
We barycentered each cleaned event list with the \texttt{\footnotesize barycen} task in order to perform precise timing studies. 
EPIC light curves were summed using the \texttt{\footnotesize LCMATH} task, with proper scaling factors for the different source photons collecting areas.
We extracted and grouped spectra with a minimum of 25 counts per bin in the $0.5-12$ keV energy band.

 \subsection{\nustar data}

The \nustar X-ray observatory was sent into orbit in the year 2012 and is notable for its exceptional sensitivity at hard X-rays. It is equipped with two X-ray grazing incidence telescopes, designated FPMA and FPMB, which are arranged in parallel and contain 2x2 solid-state CdZnTe detectors each. \nustar can operate in the energy range of 3-79 keV and can achieve an angular resolution of 18 arcsec as reported in \citet{2013ApJ...770..103H}.

\nustar observed \fgl on August 16, 2019 (58711.7031 MJD - ObsID 30502015002), with an exposure of time of ${\sim}61$~ks and on August 24, 2019 (58719.3511 MJD - ObsID 30502015004) with an exposure time of ${\sim}86$~ks. 
We processed \nustar data using {\tt NuSTARDAS-v.2.1.2} from {\tt HEASoft}~v.6.30 and {\tt CALDB} (v.20211020) calibration files. We took source events that accumulated within a circular region of 85 arcsec around the focal point. The chosen radius encloses $\sim90$\% of the point spread function. We took a circular source-free region with a radius of 160~arcsec to obtain the background events within the same CCD.

\begin{figure}[h!]
\centering
    \includegraphics[width=\columnwidth]{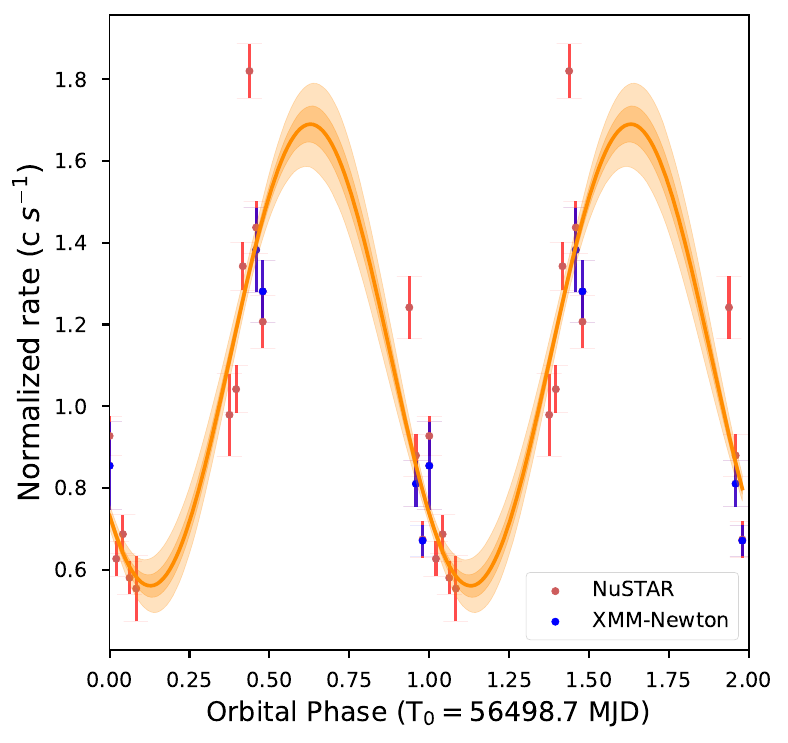}
    \caption{\nustar and \xmm folded light curve using 48 phase bins, an orbital period of 13.713 days, and with 56498.7 MJD as the reference epoch \citep{Corbet2019ApJ...884...93C}. 
    A sine function fit is shown in orange (see the main text for details). 
    The observed orbital modulation is similar to that shown by \citet{Corbet2019ApJ...884...93C} using \xrt data.}
    \label{fig:phase}
\end{figure}

We used the \texttt{\footnotesize nupipeline} task to create level 2 data products, with SAA parameters {\tt saacalc=1} {\tt saamode=strict} {\tt tentacle=yes} to filter for high-energy particle background, obtained from SAA filtering reports\footnote{\url{https://nustarsoc.caltech.edu/NuSTAR_Public/NuSTAROperationSite/SAA_Filtering/SAA_Filter.php}}. 
We extracted light curves and spectra with the \texttt{\footnotesize nuproducts} task. We obtained the barycenter-corrected light curves using \texttt{\footnotesize barycorr} task with {\tt nuCclock20100101v136} clock correction file. We used celestial coordinates ${\alpha}=211.2472^\circ$ and ${\delta}=+61.3234^\circ$ for the barycentric correction with {\tt JPL-DE200} as the Solar System ephemeris. 
Finally, we subtracted the background from each detector's light curve. Then we used the \texttt{\footnotesize LCMATH} task to create FPMA+B light curves. We extracted and re-binned spectra with a minimum of 25 counts per bin in the $3-78$ keV energy band.

We used the {\tt XSPEC}~v12.12.1 package \citep{1996ASPC..101...17A} to model \xmm and \nustar spectra, with parameters uncertainties reported at the 90\% confidence level.

\section{Results} \label{sec:results}

\subsection{Analysis of the light curves}
    
\hyperref[fig:lcnustar]{Figure~\ref{fig:lcnustar}} shows the background-corrected light curves obtained from \xmm (0.5-10 keV, top panel) and \nustar (3--78 keV, bottom panel) missions, with a bin time of 350 s. The observation conducted on August 17 is labeled as Epoch 1, while the observation on August 24 is labeled as Epoch 2. 
The long-term exposures of \nustar, both on the order of ${\sim}$1.5~d, show a hard X-ray flux modulation of ${\sim}1$~d seen in both epochs. The shorter but continuous exposures of XMM-{\sl Newton} do not capture this behavior. Instead, it captures very short (on the order of some ks) changes in soft X-ray flux, as seen in Epoch 1.

\hyperref[fig:phase]{Figure~\ref{fig:phase}} shows the orbital flux modulation associated with each observation and mission. Epoch 1 occurred within the orbital phases of 0.93--1.08, while Epoch 2 occurred within the orbital phases of 0.37--0.48. 
The observed orbital behavior of \xmm  and \nustar data is very similar to that reported by \citet{Corbet2019ApJ...884...93C} using \xrt data. 

We used a sinusoidal model with Gaussian measurement errors to visualize the orbital modulation through $\sim$10$^4$ simulations \citep{Buchner2021JOSS....6.3001B}. Specifically, the following was used:

\begin{equation}
    y = A \, \sin\left(2\pi \left[ \frac{t}{P} + t_0 \right] \right) + B + \epsilon
,\end{equation}

\noindent where $\epsilon$ $\sim$ \textsl{Normal}(0, $\sigma$), that is, a normal distribution with a mean of zero and a standard deviation of $\sigma$. We obtain the following values: $A = 0.56 \pm 0.3$, $P = 1.002 \pm 0.002$, $t_0 = 0.623 \pm 0.001$ and $B = 1.13 \pm 0.01$. The fitted model is shown in \hyperref[fig:phase]{Fig.~\ref{fig:phase}}.

From phase $\sim$ 0.93, the flux starts to decay, and from phase $\sim$ 0.37 the source has the maximum local emission. This modulation is anticorrelated with the modulation obtained from the \textit{Fermi} data in the energy range 200 MeV to 500 GeV \citep{Corbet2019ApJ...884...93C}.

We employed spectral timing routines provided by {\tt Stingray} software \citep{stingray2019ApJ...881...39H} to conduct a comprehensive search for any potential pulsation linked to the X-ray source. Light curves from both telescopes across various energy ranges do not show any significant pulsations above noise on the 0.1--100 mHz frequency range.

\begin{figure*}[h!]
\centering
    \includegraphics[width=\columnwidth]{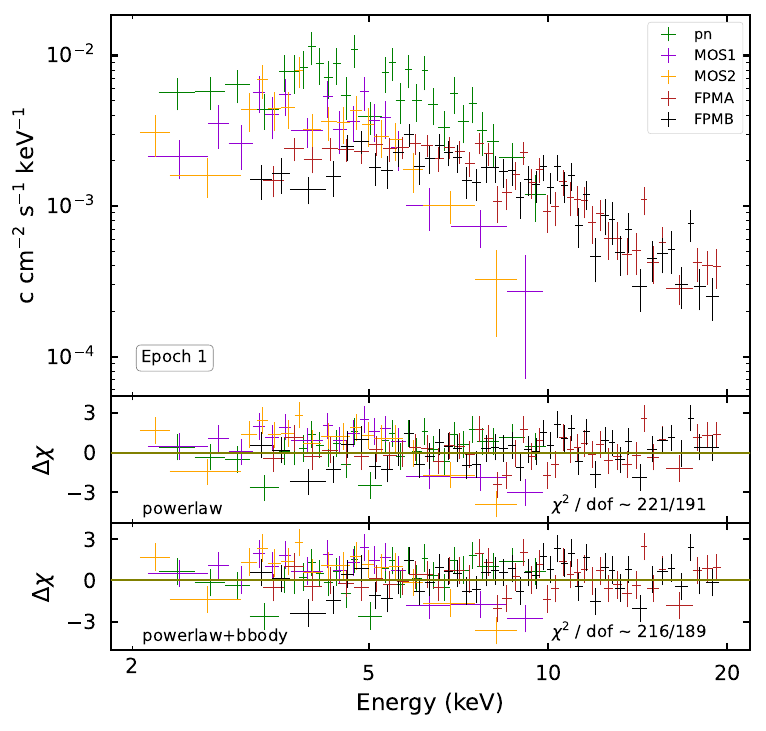}
    \includegraphics[width=\columnwidth]{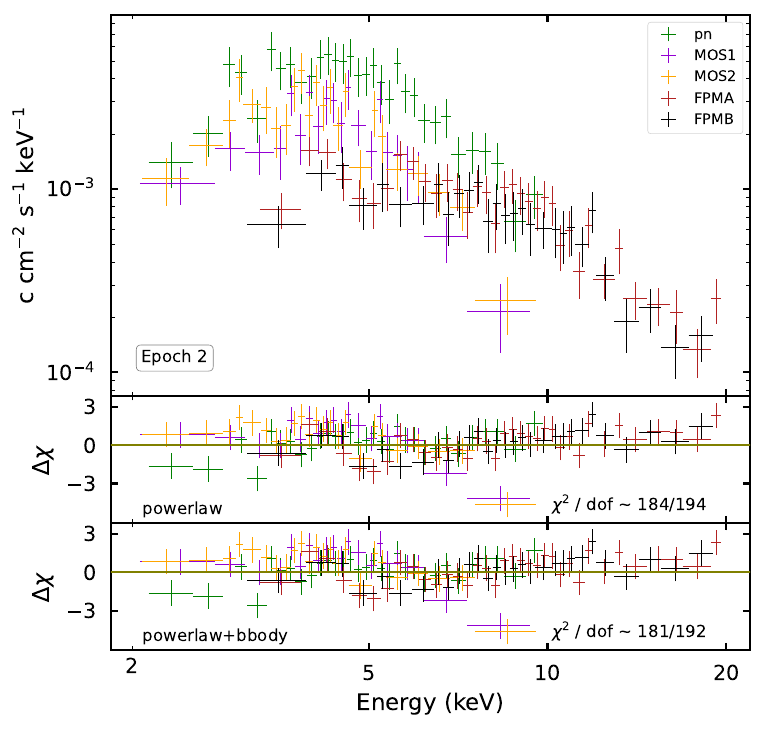}
    \caption{Spectral modeling results corresponding to {\sl Epoch 1} (left column) and {\sl Epoch 2} (right column) derived from simultaneous \xmm and \nustar data (top panels). 
    $\chi^2$ residuals correspond to an absorbed power-law (middle panel) and absorbed power-law with a black-body component (bottom panel).
 }
     \label{fig:specs}
\end{figure*}

\subsection{Spectral analysis}

We simultaneously modeled source and background spectra extracted from all five detectors (pn, MOS1, MOS2, FPMA and FPMB). We introduced calibration constants in our models to account for disparities in effective areas between instruments. The pn constant was fixed to unity, while the remaining calibration constants were permitted to vary: $C_{\rm MOS1}$ = 1.10$\pm$0.16, $C_{\rm MOS2}$ = 1.05$\pm$0.15, $C_{\rm FPMA}$=1.40$\pm$0.16, and $C_{\rm FPMB}$=1.35$\pm$0.16. Each epoch was modeled separately.

The \nustar background showed significant activity at energies above 20 keV. As a result, we limited our analysis to the energy range 3--20 keV for both epochs. In the case of the \xmm background, it was significant at energies below 2~keV. Therefore, we focused our analysis on the 2--10 keV energy range for both epochs. Consequently, the total spectrum used for each epoch included the 2--20 keV energy range.

The interstellar absorption was modeled using the Tübingen-Boulder model ({\tt tbabs}), with solar abundances defined by \citet{2000ApJ...542..914W} and effective cross sections of \citet{1996ApJ...465..487V}. 
Several continuum models were used to fit the time-averaged spectra, including combinations of power-law variants such as {\tt powerlaw}, {\tt highecut+powerlaw}, {\tt cutoffpl}, and {\tt bknpow}, and thermal models such as {\tt apec}, {\tt bbodyrad}, and {\tt diskbb}. After several fits, we selected the best fits: a power law ({\tt tbabs~powerlaw}, hereafter Model 1) and a combination of a power law and a blackbody ({\tt tbabs(powerlaw + bbodyrad)}, hereafter Model 2). No emission or absorption lines above the continuum were observed in the spectra of all epochs.

Model 1 yielded a $\chi^2$ of 221 with 191 degrees of freedom ($\chi^2_\nu=1.16$) for Epoch 1, while Epoch 2 resulted in $184/194{\sim}0.95$. On the other hand, if we apply Model 2, we obtain $216/189{\sim}1.14$ for Epoch 1 and $181/192{\sim}0.94$ for Epoch 2.

\renewcommand{\arraystretch}{1.35} 
\begin{table}[h!]
\centering
\resizebox{1\columnwidth}{!}{%
\begin{tabular}{lllll}
\hline
& \multicolumn{2}{c|}{{\sl Epoch~1}}        & \multicolumn{2}{c}{{\sl Epoch~2}} \\ \hline
\multicolumn{1}{l|}{Parameters}                         & {\tt Model 1}       & {\tt Model 2}       & {\tt Model 1}     & {\tt Model 2}    \\ \hline

\multicolumn{1}{l|}{${\rm N_H}$ [10$^{22}$~cm$^{-2}$]}  & 7.0$_{-1}^{+1.5}$      & 5$\pm$3  & 8.0$_{-1}^{+1.5}$ & $9\pm3$      \\
\multicolumn{1}{l|}{$\Gamma$}                           & 1.6$\pm$0.2 & 1.15$\pm$0.10         & 1.65$\pm$0.15   & 1.2$\pm$0.5       \\
\multicolumn{1}{l|}{Norm$_{\rm~PL}$ [$10^{-4}$]}       & 1.7$^{+0.5}_{-0.4}$  & 0.5$^{+1.5}_{-0.5}$ & 1.00$\pm$0.04      & 0.02$\pm$0.01        \\
\multicolumn{1}{l|}{kT [keV]}           & -                    & 1.0$\pm$0.3            & -               & 0.8$\pm$0.2           \\
\multicolumn{1}{l|}{$R_{\rm bb}$ [km]}                  & -        & 3$_{-2}^{+8}$      & -               & 6$_{-3}^{+10}$           \\
\multicolumn{1}{l|}{$L_x$ [10$^{33}$ {\rm erg~s$^{-1}$}]} & \multicolumn{2}{c}{$7.5\pm0.3$}   & \multicolumn{2}{c}{$4.7\pm0.3$}   \\ \hline
\multicolumn{1}{l|}{$\chi^{2}/{\rm dof}$}               & 221/191              & 216/189              & 184/194         & 181/192               \\ \hline
\end{tabular}%
}
\tablefoot{$^{\dagger}$ Parameter is fixed. Parameter confidence intervals are reported within 90\% significance. Unabsorbed luminosities, $L_x$, are calculated between 2 and 20~keV assuming a distance of 7.7 kpc.}
\caption{Best-fit parameters of \xmm+\nustar time-averaged spectral modeling of \fgl with {\tt const*tbabs(powerlaw)} (Model 1) and {\tt const*tbabs(powerlaw+bbodyrad)} (Model 2). }  
\label{tab:specs}
\end{table}

To assess the significance level of the blackbody components, we performed spectral simulations using the same observational data using {\sc fake-it} command from XSPEC. Each fake spectra was constructed from arrays of randomly sampled parameters using the {\sc simpars} command. 
By generating a cumulative distribution function of F-values from the simulated data, we determined the minimum significance level of detection by comparing it with the F-value derived from the real data, as described in \citet{Hurkett2008}.

Each F-value, for both the real and simulated data, is calculated as $F = ( \nu_0 / \delta \nu ) \times ( \delta \chi^2 / \chi^2_1)$, where the subscripts ${0,1}$ correspond to the null hypothesis ({\tt powerlaw}) and the tested hypothesis ({\tt powerlaw+bbody}). 
Each hypothesis is associated with a total $\chi^2$ and $\nu$ degrees of freedom. 
The significance level is determined by computing the corresponding $p$ value, which is the number of simulated spectra with $F$ values greater than the $F$ value derived from fitting the actual data. The uncertainty of this quantity can be calculated using $\sqrt{p(1-p)/N{\rm s}}$, where $N{\rm s}$ is the total number of simulated spectra. 

We ran $\sim$ 10$^5$ simulations for both epochs and found that the blackbody component is significant at $\sim$ 2.6$\sigma$ level for {\sl Epoch 1} and $\sim$ 2.8$\sigma$ for {\sl Epoch 2}. 

From our analysis we conclude that the spectrum was nonthermal dominated during the observed period, possibly of synchrotron origin. This implies the presence of particles with TeV energies for the typical magnetic field strengths in this type of system. In the next section we explore this hypothesis in more detail.

In \hyperref[fig:specs]{Fig.~\ref{fig:specs}} we present the time averaged spectra and residuals associated with Model 1 and Model 2 of Epoch 1 (left panel) and Epoch 2 (right panel), while the corresponding best-fit parameters and uncertainties are detailed in \hyperref[tab:specs]{Table~\ref{tab:specs}}.
The {\tt powerlaw} normalization component is expressed in units of photons~keV$^{-1}$~cm$^{-2}$~s$^{-1}$ at 1 keV. 
The {\tt bbodyrad} normalization is equal to $R_{\rm km}^2/D_{10}^2$, where $R_{\rm km}$ is the source radius in km and $D_{10}$ is the distance to the blackbody source in units of 10~kpc.
At a distance of 6.7 kpc, the size of the emitting region ranged from 1.5 to 5.4 km during Epoch 1 and from 2.9 to 8.9 km during Epoch 2. Alternatively, at a distance of 8.7 kpc, the size of the emitting region ranged from 2 to 7 km for Epoch 1 and from 4 to 11.6 km for Epoch 2. In \hyperref[tab:specs]{Table~\ref{tab:specs}}, we report the values assuming a distance of 7.7 kpc.

To compute the unabsorbed flux, we used the convolution model {\tt cflux}. Assuming a distance of 7.7~kpc \citep{Corbet2019ApJ...884...93C}, the unabsorbed 2--20 keV luminosity ranges between $7.2-7.8~\times10^{33}$~erg~s$^{-1}$ for Epoch 1 and $4.4-5~\times10^{33}$ erg~s$^{-1}$ for Epoch 2.


\section{Model}
\label{sec:model}

\subsection{Jet nonthermal radiation}

We adopted the hypothesis that a jet is present in the $\gamma$-ray-emitting binary \fgl, and tried to evaluate whether it can adequately explain the nonthermal spectral energy distribution (SED) of the system. The radiative jet model used follows that presented in detail in \citet{Escobar2022}, which in turn is based on \citet{Romero2008}. In the following, the model setup is outlined.

 The scenario consists of a CO accreting material from the companion star with an accretion power $L_{\mathrm{acc}}$, which can be expressed in terms of the Eddington luminosity as

 \begin{eqnarray}
L_{\mathrm{acc}} = q~L_{\mathrm{Edd}} \approx q~1.3 \times 10^{38} \left(\frac{M}{M_{\odot}}\right)~\mathrm{erg}~\mathrm{s}^{-1},
\end{eqnarray}

\noindent where $q$ is a constant that represents the accretion efficiency in terms of the Eddington limit. Coupled with the inner accretion disk we assume the presence of a lepto-hadronic jet of kinetic luminosity, $L_{\mathrm{jet}}$. This jet power relates to the accretion power through
 
\begin{eqnarray}
L_{\mathrm{jet}} = q_{\mathrm{jet}} L_{\mathrm{acc}},
\end{eqnarray}

 \noindent where $q_{\mathrm{jet}}$ is another constant indicating the fraction of the accretion power that is transferred to the jet. The parameters $q$ and $q_{\mathrm{jet}}$ define the accretion regime of the MQ. To compute the SED and the normalizations with the emission power, we chose to use $L_{\mathrm{jet}}$ directly as the parameter instead of a combination of $q$ and $q_{\mathrm{jet}}$; we defer discussion of the interpretation of the accretion regime to \hyperref[sec:disc]{Sect.~\ref{sec:disc}}.
 
 The jet propagates with a bulk velocity $v_{\mathrm{jet}}$, corresponding to a bulk Lorentz factor $\Gamma_{\mathrm{jet}}$. A fraction $q_{\mathrm{rel}}$ of this power is converted into relativistic particles by an acceleration mechanism. The relativistic proton and electron luminosities, $L_{\mathrm{p}}$ and $L_{\mathrm{e}}$, are distributed according to the power ratio $a = L_{\mathrm{p}} / L_{\mathrm{e}}$.

We use cylindrical coordinates, with the coordinate $z$ along the jet axis and the origin at the CO. The jet is started at a distance $z_{0}$ from the CO. The region in which the particles are accelerated extends from $z_{\mathrm{min}}$ to $z_{\mathrm{max}}$, and its shape is described by

\begin{eqnarray}
r(z) = r_{0} \left( \frac{z}{z_{0}} \right)^{\varepsilon},
\end{eqnarray}
 
\begin{figure*}[t]
\centering
    \includegraphics[width=0.8\textwidth]{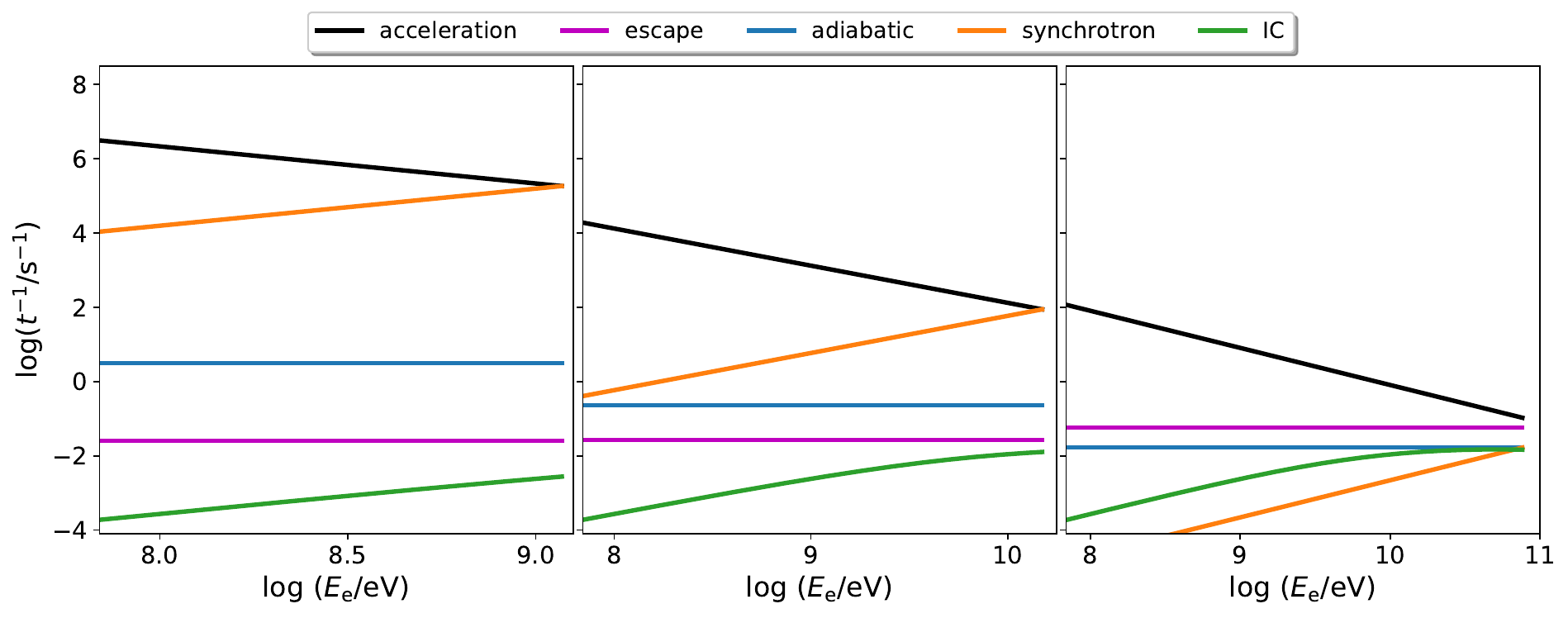} 
    \includegraphics[width=0.8\textwidth]{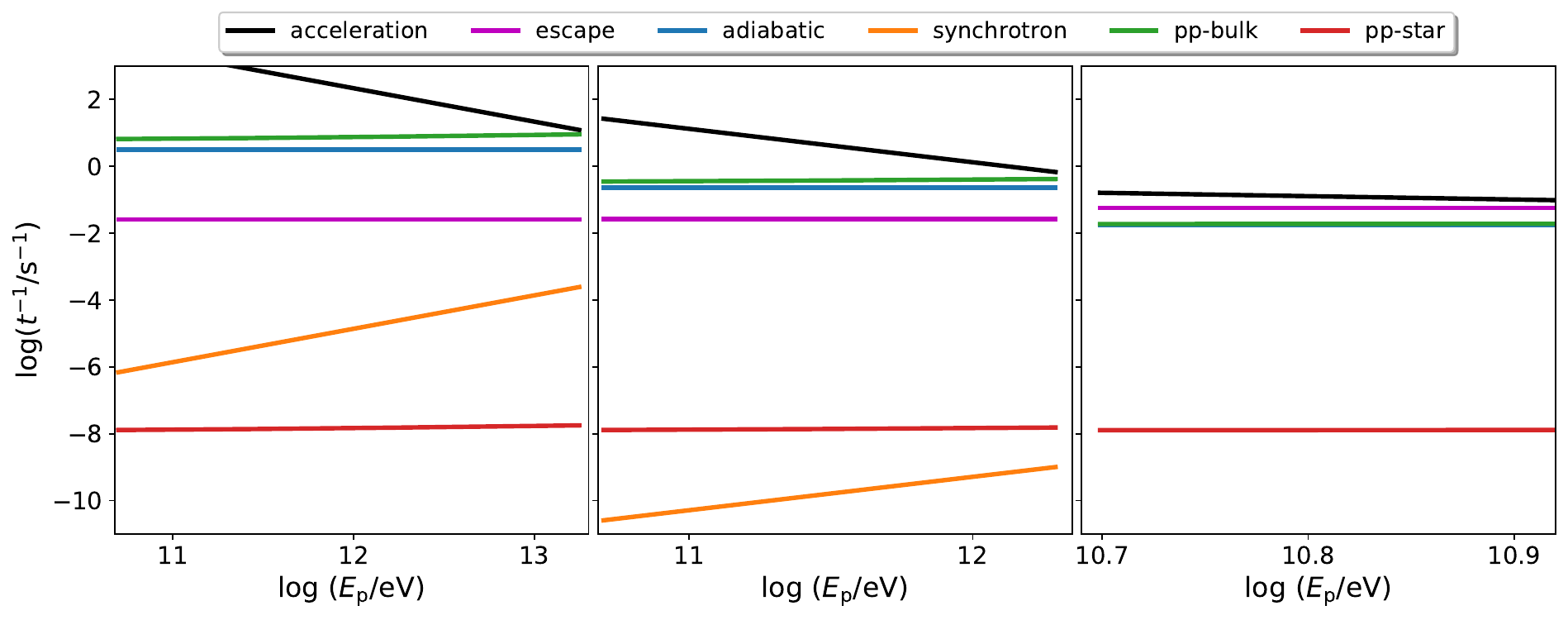}
    \caption{Acceleration, escape, and cooling rates of electrons (top row) and protons (bottom row). Each column shows the aforementioned rates calculated at $z_{\rm min}$ (left column), at the logarithmic midpoint (middle column), and at $z_{\rm max}$ (right column).}
    \label{fig:ratescool}
\end{figure*}

  \begin{figure}[t]
\centering
    \includegraphics[width=\columnwidth]{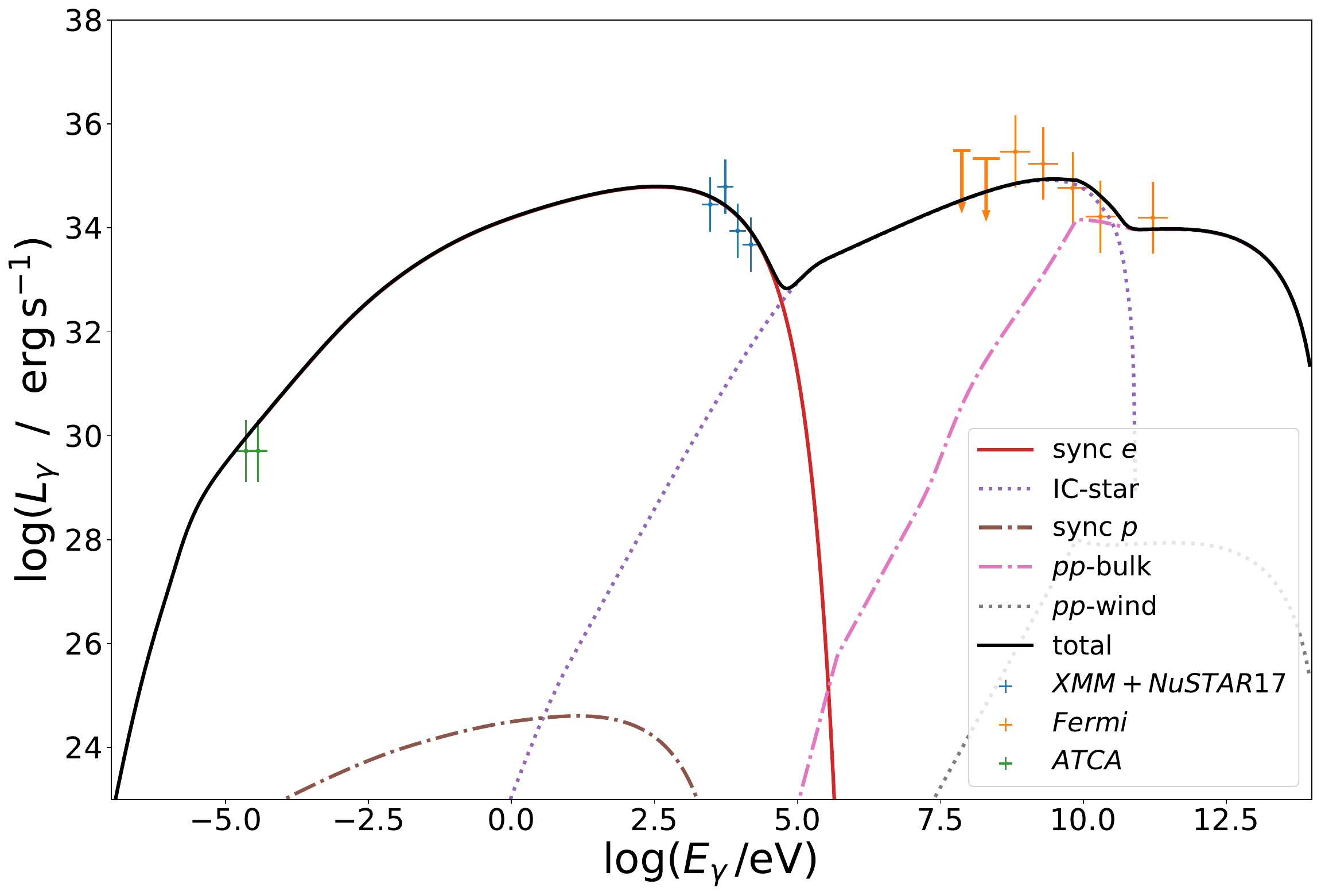}
    \caption{Nonthermal SED derived from our jet model. We considered the following radiative processes: for protons, synchrotron emission and proton-proton collisions with the cold protons of the jet bulk and the companion's wind; for electrons, synchrotron emission and inverse Compton scattering off the radiation field of the companion. 
    The figure also shows luminosities derived from \xmm+\nustar data (our work) as well as ATCA and {\it Fermi} data taken from \citet{Corbet2019ApJ...884...93C} and the 4FGL-DR2 catalog \citep{Abdollahi2020ApJS..247...33A, Ballet2020arXiv200511208B}, respectively. All the references are in the figure.}
    \label{fig:sed}
\end{figure}

\noindent where $r_{0}$ is the radius of the jet at its base and $0 < \varepsilon \leq 1$ describes its geometry. We note that the degree of collimation of the jet increases with decreasing values of $\varepsilon$, with the extreme value $\varepsilon = 1$ representing a conical shape. The magnetic field along the jet, $B$, decreases with $z$ following a power law of index $m$, namely,  $B(z) = B_{0} (z / z_{0})^{-m}$, where $1 \leq m \leq 2$ \citep[e.g.,][]{Krolik1999}. The value of $B_{0}$ is obtained by assuming equipartition between magnetic and kinetic energy at the jet base. 

We parameterize the injection function of energy to relativistic particles as a power law with an exponential cutoff,

\begin{eqnarray}
Q_{i} = Q_{i,0}E_{i}^{-p} \exp \left( \frac{E}{E_{i,\mathrm{max}}} \right),
\end{eqnarray}

\noindent where $i = \mathrm{e}, \mathrm{p}$ accounts for electrons and protons, respectively; $Q_{i,0}$ is obtained normalizing the injection function with the total power of each particle population, $L_{i}$; and $E_{i,\mathrm{max}}$ is the maximum reachable energy, achieved when the acceleration rate equals that of energy losses.

Relativistic particles are accelerated at a rate of

\begin{eqnarray}
t^{-1}_{\mathrm{acc}} =\frac{\eta c e B}{E_{i}},
\end{eqnarray}

\noindent where $\eta \leq 1$ is the acceleration efficiency, $c$ is the speed of light, and $e$ is the elementary electric charge. On the other hand, relativistic particles lose energy via both radiative and non-radiative mechanisms. The latter are adiabatic and escape losses for both proton and electron populations. Regarding the radiative mechanisms, we consider synchrotron emission for both types of particles. In the case of protons, we also considered photons from the decay of neutral pions, which are the product of inelastic collisions between relativistic and cold protons; the latter consist mainly of protons in the jet bulk and those in the stellar wind of the companion star. 
In the case of electrons, we also computed the inverse Compton scattering of photons from the stellar radiation. 

We assume that the relativistic particle populations reach a steady state. Their spectral densities are obtained by solving the steady state transport equation, taking injection, escape, and continuous losses into account. 
For a discussion of the general form of the transport equation, we refer the reader to \citet{Ginzburg1964}. 

The complete formulae for calculating of radiation processes in this work can be found in \citet{Blumenthal1970}, \citet{Begelman1990}, \citet{Atoyan2003}, \citet{Kafexhiu2014}, and references therein.


\subsection{Spectral energy distribution of \fgl}

 To fit the SED of the source with our model, we used the radio observations of \citet{Corbet2019ApJ...884...93C}, the X-ray flux obtained in this work, and the $\gamma$-ray flux from the 4FGL-DR2 catalog \citep{Abdollahi2020ApJS..247...33A, Ballet2020arXiv200511208B}.
 
In \hyperref[fig:ratescool]{Fig.~\ref{fig:ratescool}} we show the acceleration, escape, and cooling rates of the relativistic particles at different locations in the jet. We find that for the electron population, the losses are dominated by synchrotron cooling throughout the emitting region. For the protons, both adiabatic and proton-proton collisions with bulk protons are the dominant mechanisms of energy loss in the region close to the base; the escape rate competes with them in regions further from the base and dominates the losses toward the end of the emitting region.

In \hyperref[fig:sed]{Fig.~\ref{fig:sed}} we show the derived nonthermal SED of the jet, $L_{\gamma}$, which includes all the above radiative processes, where 

\begin{eqnarray}
L_{\gamma} = E_{\gamma}^{2} \frac{dN}{dE_{\gamma}dt},
\end{eqnarray}

\noindent and $dN$ is the number of photons with energies between $(E_{\gamma}, E_{\gamma}+dE_{\gamma})$ emitted during a time $dt$. The assumed and derived parameters of the model, with uncertainties reported at the 90\% confidence level for the free parameters, are listed in \hyperref[tab:params]{Table~\ref{tab:params}}. 
The X-ray data points shown in \hyperref[fig:sed]{Fig.~\ref{fig:sed}} correspond to Epoch 1. 
Since there is no significant change in the X-ray flux between the two epochs (see \hyperref[fig:lcnustar]{Fig.~\ref{fig:lcnustar}}), the same set of parameters also fits the observations of Epoch 2.

To obtain the reported model parameters, we ran a first set of ${\sim}100$ simulations\footnote{We use the term "simulation" to mean the computation of the total SED with a defined set of parameters, i.e., a running of our model for a specific implementation of the parameters.} covering a wide range of values, and decided which of these remained fixed, apart from those resulting from observed properties of the system. Then we ran another set of $120$ simulations considering the variation of all free parameters (i.e., using $40$ simulations for each parameter variation at a time; $m$, $p$, $\varepsilon$, and $\eta$), and chose the model with the minimum $\chi^{2}/{\rm dof}$. The set of values of the free parameter space was covered with a uniform grid for $p$, $\varepsilon$, $m$, and $\log\eta$. To estimate the free parameter errors, we computed the chi-squared for each simulation and chose the $\chi_{\mathrm{min}}^{2} + 2.706$ contour for each free parameter, which represents the $90~\%$ credible region \citep[e.g.,][]{Frodesen1979}. The errors are reported in \hyperref[tab:params]{Table~\ref{tab:params}}. In the particular case of $m$ ($\eta$) the upper (lower) bound on the error comes from restricting the possible values to those reported in the fourth column of \hyperref[tab:params]{Table~\ref{tab:params}}, while for the case of $\varepsilon$ all the values within the bounds fall below the aforementioned contour. We obtained a minimum chi-squared goodness of fit of $\chi^{2}/$d.o.f$ = 1.06$, with $7$ d.o.f.

We show that the broad spectrum of \fgl can be explained by the nonthermal emission associated with a mildly relativistic ($\Gamma_{\mathrm{jet}} = 1.9$), lepto-hadronic model of a MQ jet. In particular, this model represents an approximately parabolic jet ($\varepsilon \approx 0.56$), with a compact emitting region of size $\approx 1.0 \times 10^{12}~\mathrm{cm}$ (jet extension could be orders of magnitude larger), and a magnetic induction field at its base of $B_{0} \approx 2.8\times 10^{7}~\mathrm{G}$. 
Relativistic protons and electrons share the total power in the relativistic populations, $L_{\mathrm{rel}} = q_{\mathrm{rel}} L_{\mathrm{jet}} = 10^{37}~\mathrm{erg}~\mathrm{s}^{-1}$, according to an assumed hadron-to-lepton ratio of $a \approx 0.11$, and are accelerated through a low-efficiency mechanism for which $\eta \approx 1.0 \times 10^{-4}$. 
The particle injection function shows a hard spectrum with spectral index $p = 1.98$. 

\begin{table}[ht!]
        \centering
        \resizebox{\columnwidth}{!}{%
        \begin{tabular}{lllll}
        \hline
        \hline
        \!\!\!\!Parameter & \!\!\!Symbol & \!\!\!\!Adopted/Derived Value & Typical Values & Units\\
        \hline
        \!\!\!\!Luminosity & \!\!\!$L_{\mathrm{jet}}$ & \!\!\!\!$10^{38}$ & $10^{36}-10^{39}$ & $\mathrm{erg\,s^{-1}}$ \\
        \!\!\!\!Launching distance & \!\!\!$z_{0}$ & \!\!\!\!$1.0\times10^{8}$ & $\gtrsim 10^{8}$ & $\mathrm{cm}$\\
        \!\!\!\!Magnetic field at $z_{0}$ & \!\!\!$B_{0}$ & \!\!\!\!$2.8\times10^{7}$ & $\sim 10^{5}$ -- $10^{7}$ & $\mathrm{G}$\\
        \!\!\!\!Base of acceleration region & \!\!\!$z_{\mathrm{a}}$ & \!\!\!\!$3.0\times10^{9}$ & $\gtrsim 10^{8}$ & $\mathrm{cm}$\\
        \!\!\!\!Top of acceleration region & \!\!\!$z_{\mathrm{t}}$ & \!\!\!\!$1.0\times10^{12} $ & up to jet extent & $\mathrm{cm}$\\
        \!\!\!\!Base half-opening angle$^{\dagger}$ & \!\!\!$\theta_{\mathrm{jet}}$ & \!\!\!\!$0.3$ & $\lesssim 10$ & ${\mathrm{deg}}$\\
         \!\!\!\!Viewing angle & \!\!\!$\theta$ & \!\!\!\!$60$ & $0 - 180$ & ${\mathrm{deg}}$\\
        \!\!\!\!Bulk Lorentz factor & \!\!\!$\Gamma_{\mathrm{jet}}$ & \!\!\!\!$1.9$ & $\sim1-5$ &\\
        \!\!\!\!Relativistic power fraction & \!\!\!$q_{\mathrm{rel}}$ & \!\!\!\!$0.1$ & $\lesssim 0.1$ &\\
        \!\!\!\!Proton-electron power ratio & \!\!\!$a$ & \!\!\!\!$0.11$ & $0-100$ & \\
        \!\!\!\!Magnetic power-law index & \!\!\!$m$ & \!\!\!\!$1.95\substack{+0.05 \\ -0.08}$ & $1-2$ & \\
        \!\!\!\!Acceleration efficiency log & \!\!\!$\log\eta$ & \!\!\!\!\!\!\!\!$-4.01\substack{+0.26 \\ -0.99}$ & $-5 \leq \log\eta \leq -1$ & \\
    \vspace{-0.31cm} \\
        \!\!\!\!Injection spectral index & \!\!\!$p$ & \!\!\!\!$1.98\substack{+0.35 \\ -0.29}$ & $1.5-2.5$ & \\ \vspace{-0.31cm} \\
        \!\!\!\!Geometric index & \!\!\!$\varepsilon$ & \!\!\!\!$0.56\substack{+0.44 \\ -0.46}$ & $0.1 \leq \varepsilon \leq 1$ & \\

        \hline
        \end{tabular}%
        }
        \tablefoot{$^{\dagger}\, \tan\theta_{\mathrm{jet}} = r_{0} / z_{0}$.}
        \caption{Jet model parameters.}
   \label{tab:params}
        \end{table}

In all cases, the values of the parameters are consistent with their commonly assumed range of values for MQ jets \citep[see, for example,][]{Romero2008, Vila2012, Pepe2015, Escobar2021}. The SED is computed assuming a viewing angle of $\theta = 30^{\circ}$ (i.e., the angle between the jet axis and the line of sight). With the Lorentz factor from \hyperref[tab:params]{Table~\ref{tab:params}}, and for lower viewing angles, the same data could be explained with a less powerful jet ($\sim 0.07$ times the one reported here), while higher angles would favor a scenario with more powerful jets, up to a factor of $\sim 50$. Instead, maintaining fixed all the values of the parameters but the viewing angle, the model still manages to explain the observations with at most a variation of $\approx 10^{\circ}$. As we can see, knowing the viewing angle is a crucial factor in determining the accretion regime of the emitter.


\section{Discussion and conclusions} \label{sec:disc}

In this work we have analyzed simultaneous \xmm (2--10 keV) and \nustar (3--20 keV) observations of the $\gamma$-ray-emitting binary \fgl.
The difference between the two observed epochs is about six days. We found no significant pulsation above statistical noise in the \xmm and \nustar light curves. This non-detection leaves open the question of the nature of the CO. 

We fitted the time-averaged spectra of both epochs with empirical models, which yielded continuum parameters consistent with $\gamma$-ray-emitting binaries \citep{Kretschmar2019NewAR..8601546K, An2015ApJ...806..166A, Yoneda2021ApJ...917...90Y} and MQs \citep{Natalucci2014ApJ...780...63N, Soria2020ApJ...888..103S, Hirsch2020A&A...636A..51H, Rodi2021ApJ...910...21R, saavedra2022BAAA...63..274S}. 
For Epoch 1, a power-law model adequately characterized the spectrum. For Epoch 2, however, the addition of a blackbody component significantly improved the fit to the observed data. The power-law index was found to be ${\sim}1.7$ for Epoch 1 and ${\sim}1.4$ for Epoch 2. Furthermore, the associated blackbody temperature for Epoch 2 was ${\sim}0.8$ keV, corresponding to a compact region of radius ${\lesssim}~16$ km tentatively associated with the inner region of an accretion disk. 

\citet{Corbet2019ApJ...884...93C} present a different explanation for the emission of \fgl. They set out a scenario involving the collision of winds between a rapidly rotating NS and its stellar companion. In addition, \citet{Xingxing2020} modeled this scenario to explain the GeV emission from \fgl. In the colliding-wind scenario, the interaction between the stellar and pulsar winds gives rise to shocked regions that are characterized by a spiral shape due to Coriolis forces \citep[see][]{Molina2020}. Within this framework, the high-energy emission in a $\gamma$-ray binary system can be attributed to the up-scattering of photons from the stellar radiation field due to inverse Compton scattering by the relativistic electrons and positrons present in the shocked fluid. Synchrotron emission contributes to a lesser extent, especially in the energy range around ${\sim}1$ MeV \citep{Molina2020}.

To explain the new X-ray data and archival multiwavelength observations (ATCA and \textit{Fermi}), we implemented a lepto-hadronic jet model under the hypothesis that \fgl is a MQ. The parameters associated with the properties of the relativistic particle populations in the jet and the nonthermal emission are consistent with the values commonly assumed for this type of source \citep[see, for example,][]{Romero2008, Vila2012, Pepe2015, Escobar2021, Escobar2022}. The scenario consists of a parabolic jet with a compact acceleration region, where the high-energy emission is produced by a hard spectrum of relativistic particles driven by a low-efficiency acceleration mechanism.

In this scenario, the $\gamma$-ray emission is produced via two processes. First, inverse Compton scattering produces nonthermal radiation in the energy range from about 100 keV to 10 GeV. Second, the decay of neutral pions from proton-proton collisions produces energetic photons at energies above 10 GeV. 
At lower energies, the synchrotron emission of electrons completely dominates, allowing us to accurately model radio and X-ray data. We thus propose that a lepto-hadronic jet model, which includes both leptonic and hadronic processes, may be sufficient to explain the multiwavelength emission of \fgl.

 There are essentially two interpretations of the results we obtained with our jet model. 
 If the accretion regime is sub-Eddington, the CO should be a BH of at least ${\sim}10~M_{\odot}$. This scenario, like the case of \object{Cygnus X-1}, would require an X-ray luminosity on the order of $\sim 10^{37}~\mathrm{erg}~\mathrm{s}^{-1}$ in the low-hard state \citep{BeppoSax2001, SuzakuCgnX12008}, which is in apparent contradiction with the observations. 
 On the other hand, if the source is in a super-Eddington accretion regime, this would imply that the CO consists of a BH of a few solar masses or a NS with a weak magnetic field. In this case, the X-ray emission from the disk is absorbed by the photosphere of the wind ejected by the same disk \citep{Abaroa2023}. This is more consistent with what is observed in the X-ray emission. The supercritical source may also have an equatorial radio component or equatorial lines with velocities of $10^3-10^4$ km s$^{-1}$, as in the case of SS 433, which could be observed in the future \citep{Fabrika2021AstBu..76....6F,Abaroa2023}.

We note that leptonic jet models can also be adopted to explain observations of other MQs, as in the cases of \object{Cyg X-3} \citep{Zdziarski2012} and \object{Cyg X-1} \citep{Zdziarski2014}. In our case, however, this type of model would not fully explain the observations for two main reasons. First, there is no clear evidence for disk or coronal emission. The X-ray observations could then be explained by a dominant synchrotron emission, which at these energies hides the radiation from the disk and/or the corona \citep[see, for example, Fig. 11 of ][]{Bosch-Ramon2006}. On the other hand, a relativistic proton component seems necessary to explain the observed slope change in the high-energy part of the \fermi data. In the lepto-hadronic picture, this part of the spectrum is dominated by the $\gamma$-ray emission, which originates in proton-proton collisions through neutral pion decays.
 
 As shown in \hyperref[fig:phase]{Fig.~\ref{fig:phase}}, the flux is higher in Epoch 1 than in Epoch 2. This behavior is also consistent with the presence of a thermal component in the spectra of Epoch 2, which can be explained by the contribution of a thermal inner disk and a reduced nonthermal jet component. In the case of a moderate or low viewing angle, some of the emission from the disk can escape from the central funnel in the wind \citep{Abaroa2023}. As for Epoch 1, the contribution to the total flux may be completely dominated by the jet.
 
 The method we implemented for estimating parameters and uncertainties, although not very robust, allowed us to explain the multiwavelength behavior of \fgl, favoring the MQ scenario.
 A more detailed analysis using a Markov chain Monte Carlo method would improve these estimates. 
 This method would also allow a model comparison test to be included (e.g., to compute the odds ratio between leptonic and lepto-hadronic models, or between the MQ and colliding-wind scenarios). This analysis is beyond the scope of the current work, but we will present it in a future paper.

It is important to collect more observational data to evaluate the pros and cons of the pulsar-star collision wind and MQ scenarios for \fgl. These additional data should be used to focus on timing analysis, especially with respect to the orbital period. By analyzing these data, we can gain a more complete understanding of the nature of this and other $\gamma$-ray-emitting binary systems.

\begin{acknowledgements}

We thank the anonymous reviewer for their valuable comments on this manuscript. We extend our gratitude to Fiona A. Harrison and Brian W. Grefenstette for their assistance with the instruments' technical aspects. We thank Sergio Campana for his suggestions that helped improve this work. EAS, FAF and JAC acknowledge support by PIP 0113 (CONICET). FAF is fellow of CONICET. JAC is CONICET researchers. JAC is a Mar\'ia Zambrano researcher fellow funded by the European Union  -NextGenerationEU- (UJAR02MZ). This work received financial support from PICT-2017-2865 (ANPCyT). JAC was also supported by grant PID2019-105510GB-C32/AEI/10.13039/501100011033 from the Agencia Estatal de Investigaci\'on of the Spanish Ministerio de Ciencia, Innovaci\'on y Universidades, and by Consejer\'{\i}a de Econom\'{\i}a, Innovaci\'on, Ciencia y Empleo of Junta de Andaluc\'{\i}a as research group FQM-322, as well as FEDER funds. GER acknowledges financial support from the State Agency for Research of the Spanish Ministry of Science and Innovation
under grants PID2019-105510GB-C31AEI/10.13039/501100011033/ and
PID2022-136828NB-C41/AEI/10.13039/501100011033/, and by “ERDF A way of
making Europe”, by the "European Union", and through the "Unit of
Excellence Mar\'ia de Maeztu 2020-2023" award to the Institute of
Cosmos Sciences (CEX2019-000918-M). Additional support came from PIP 0554 (CONICET). GJE acknowledges financial support from the European Research Council for the ERC Consolidator grant DEMOBLACK, under contract no. 770017.

\end{acknowledgements}

%
%
\bibliographystyle{aa}
\bibliography{biblio}
\end{document}